\documentclass[12pt]{article}
\usepackage{amsmath, amssymb, enumerate, amsthm}
\usepackage{bm}
\usepackage{graphicx}
\usepackage{booktabs}
\usepackage{siunitx}
\usepackage{hyperref}
\usepackage{natbib}

\DeclareMathOperator*{\argmin}{arg\,min}
\DeclareMathOperator{\diag}{diag}

\DeclareMathOperator*{\vectorize}{vec}

\newtheorem{Algorithm}{Algorithm}

\theoremstyle{plain}

\newtheorem{Theorem}{Theorem}
\newtheorem{Remark}{Remark}

\newtheorem{Corollary}{Corollary}

\newcommand{\blind}{1}

\begin{document}

\def\spacingset#1{\renewcommand{\baselinestretch}%
{#1}\small\normalsize} \spacingset{1}


\if1\blind
{
  \title{\bf Divisive Hierarchical Clustering Using Block Diagonal Matrix Approximations}
  \author{Jan O. Bauer\\
    Department of Econometrics and Data Science, Vrije Universiteit Amsterdam, \\
    Amsterdam, The Netherlands\\
    Tinbergen Institute, Amsterdam, The Netherlands}
  \maketitle
} \fi

\if0\blind
{
  \bigskip
  \bigskip
  \bigskip
  \begin{center}
    {\LARGE\bf Divisive Hierarchical Clustering Using Block Diagonal Matrix Approximations}
\end{center}
  \medskip
} \fi

\bigskip
\begin{abstract}
\noindent
In this work, we introduce a novel methodology for divisive hierarchical clustering. Our divisive (``top-down'') approach is motivated by the fact that agglomerative hierarchical clustering (``bottom-up''), which is commonly used for hierarchical clustering, is not the best choice for all settings. The proposed methodology approximates the similarity matrix by a block diagonal matrix to identify clusters. While divisively clustering $p$ elements involves evaluating $2^{p-1}-1$ possible splits, which makes the task computationally costly, this approximation effectively reduces this number to at most $p(p-1)$ candidates, ensuring computational feasibility. We elaborate on the methodology and describe the incorporation of linkage functions to assess distances between clusters. We further show that these distances are ultrametric, ensuring that the resulting hierarchical cluster structure can be uniquely represented by a dendrogram, with interpretable heights. Additionally, the proposed methodology exhibits the flexibility to also optimize objectives of other clustering methods, and it can outperform these. The methodology is also applicable for constructing balanced clusters. To validate the efficiency of our approach, we conduct simulation studies and analyze real-world data. Supplementary materials for this article can be accessed online.
\end{abstract}

\noindent%
{\it Keywords:}  Balanced clustering; Clustering; Hierarchical clustering; Matrix approximation; Singular value decomposition; Variable clustering
\vfill

\newpage
\spacingset{1.9} 
\section{Introduction}
\label{s:Introduction}

Cluster analysis aims to find structure in data by identifying groups that are similar in some sense and is a common step in the exploratory analysis of data. As an unsupervised learning technique, clustering does not rely on pre-specified class labels. Broadly, clustering methods can be divided into two categories: partitional and hierarchical. Partitional clustering divides data into a pre-determined number of clusters, while hierarchical clustering produces a cluster tree (dendrogram), which is a tree-like representation that records the successive merging or splitting of clusters. The dendrogram provides a flexible view of the clustering structure, allowing the process of merging and splitting to be examined at different levels.

In general, there are two main approaches for hierarchical clustering: divisive (``top-down'') and agglomerative (``bottom-up'') clustering. Divisive clustering starts with a single cluster containing all elements (\textit{top}) and iteratively splits one cluster into two, until each element represents one cluster (\textit{bottom}). In agglomerative clustering, on the contrary, each element is considered as one cluster (\textit{bottom}) and the elements are clustered iteratively until all elements are contained in a single cluster (\textit{top}). Both approaches group the data by creating a dendrogram based on specific distances between points and similarity or dissimilarity measures \citep{HTF09, RO18}.

\begin{figure}[!htbp]
\center
\includegraphics[width=.9\textwidth]{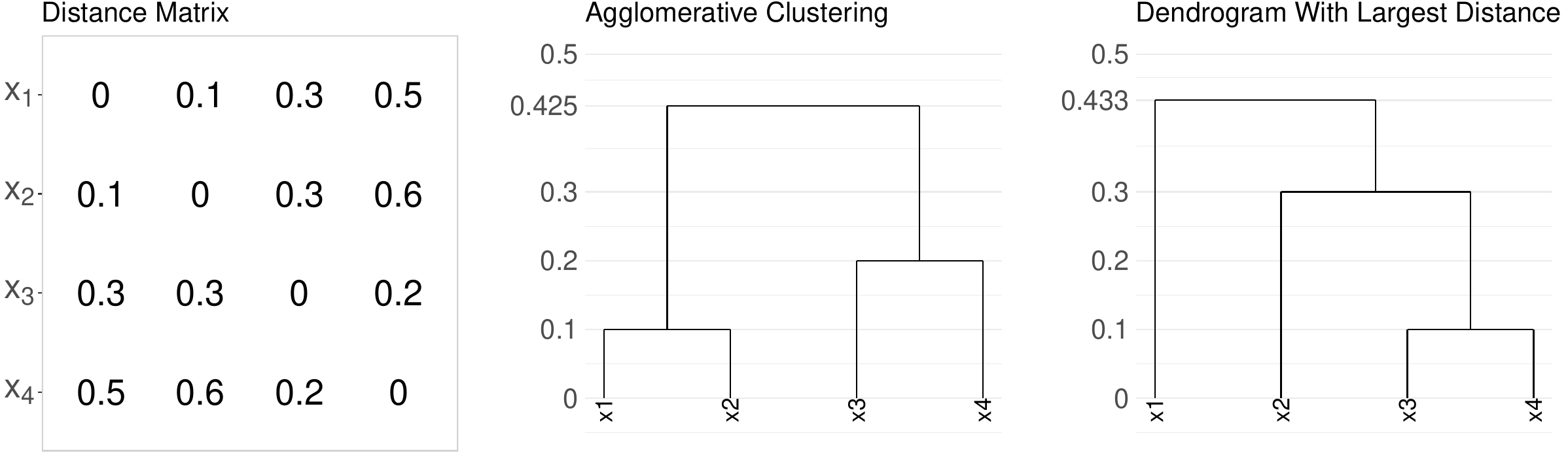}
  \caption{Example of a pairwise distance matrix (\textit{left}) where agglomerative hierarchical clustering using average linkage fails to detect the largest distance for the hierarchical structure (\textit{center}). In this case, agglomerative clustering merges $\{x_1\}$ with $\{x_2\}$, and $\{x_3\}$ with $\{x_4\}$ in the initial steps, resulting in a final merge of $\{x_1, x_2\}$ and $\{x_3, x_4\}$ with a distance of $d(\{x_1, x_2\}, \{x_3, x_4\})  = 0.425$. However, this is not the largest possible distance, as the clusters $\{x_1, x_2, x_3\}$ and $\{x_4\}$ exhibit the distance $d(\{x_1, x_2, x_3\}, \{x_4\}) \approx 0.433 > 0.425$ between them. The hierarchical structure representing this largest distance is shown on the \textit{right}.}
  \label{fig:IntroductionMisspec}
\end{figure}

However, agglomerative methods are not always the optimal choice. While they are well-suited for grouping elements into many small clusters, they become less ideal when grouping elements into only a few clusters, which is often the primary goal in clustering analysis. This limitation arises from the greedy nature of agglomerative methods, which can lead to suboptimal clusters when the algorithm proceeds to the top of the hierarchy. To illustrate this, we consider a simple example involving four objects: $x_1$, $x_2$, $x_3$, and $x_4$, which we aim to cluster using average linkage. The pairwise distance matrix for this example is given in Figure~\ref{fig:IntroductionMisspec} (\textit{left}). In agglomerative clustering, each variable is initially treated as a separate cluster, and the two clusters with the smallest average distance are merged iteratively. In the first step, $\{x_1\}$ and $\{x_2\}$ are merged into the cluster $\{x_1, x_2\}$ because the average distance $d(\{x_1\}, \{x_2\}) = 0.1$ between these two clusters is the smallest among all clusters. In the second step, $\{x_3\}$ and $\{x_4\}$ are merged into the cluster $\{x_3, x_4\}$ with distance $d(\{x_3\}, \{x_4\}) = 0.2$. Finally, the clusters $\{x_1, x_2\}$ and $\{x_3, x_4\}$ are merged with a distance of $d(\{x_1, x_2\}, \{x_3, x_4\}) = (0.3 + 0.5 + 0.3 + 0.6)/4 = 0.425$ between them. The resulting dendrogram is given in Figure~\ref{fig:IntroductionMisspec} (\textit{center}), indicating that the two clusters $\{x_1, x_2\}$ and $\{x_3, x_4\}$ have the largest distance between them. However, this is a misspecification, as the largest possible distance is between the clusters $\{x_1, x_2, x_3\}$ and $\{x_4\}$ with $d(\{x_1, x_2, x_3\}, \{x_4\}) = (0.5 + 0.6 + 0.2)/3 \approx 0.433 > 0.425$. The hierarchical structure that correctly captures this distance is shown in Figure~\ref{fig:IntroductionMisspec} (\textit{right}) and can be identified by calculating the distances for all possible splits of $\{x_1, x_2, x_3, x_4\}$ into two clusters, as performed in divisive clustering. The agglomerative approach, in this case, fails to uncover this structure.

Therefore, divisive clustering alternatives are needed. However, when divisively clustering $p$ objects, there are $2^{p-1}-1$ candidate splits which is computationally demanding. \citet{KR90} introduced divisive analysis clustering (DIANA). DIANA selects the object with the highest average dissimilarity and forms a new cluster with it. It then moves all objects that are more similar to this new cluster than to the remaining objects into the newly formed cluster. Consequently, not all $2^{p-1}-1$ candidate splits need to be examined.  \citet{BD08, BD10} developed the average correlation coefficient algorithm (ACCA) based on a similarity (correlation) matrix, which moves objects from one cluster to another if this increases the average correlation of the first cluster, and provided that the object does not reduce the average correlation of the cluster it is moved to. \citet{CVZ20, CVZ24} identified hierarchical clusters by finding a so-called ultrametric correlation matrix that fits the underlying data. This methodology, abbreviated as UCOR, is neither agglomerative nor divisive, as it aims to determine the hierarchical structure directly through the ultrametric correlation matrix.

\begin{figure}[!htbp]
\center
\includegraphics[width=.55\textwidth]{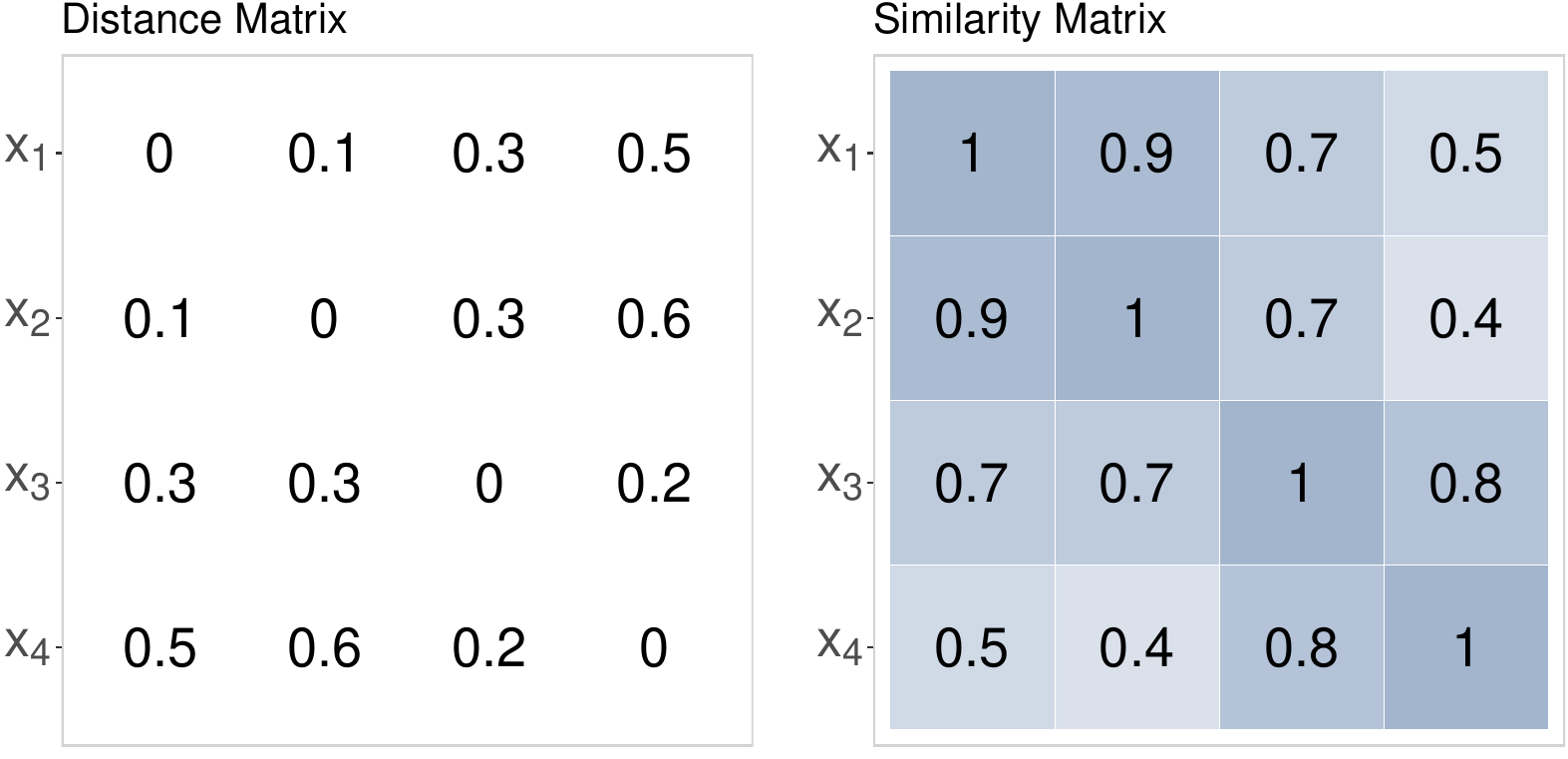}
  \caption{Distance matrix $\bm{D} = \bm{1} - \bm{S}$ from the example corresponding to Figure~\ref{fig:IntroductionMisspec} (\textit{left}) with corresponding similarity matrix $\bm{S}$ (\textit{right}).}
  \label{fig:Introduction2}
\end{figure}

As we will demonstrate later, ACCA and UCOR have long runtimes, while ACCA, DIANA and UCOR have difficulties in identifying all types of nested structures, which does not make them competitive alternatives to agglomerative clustering. To the best of our knowledge, no substantial efforts have been made in the direction of non-agglomerative hierarchical clustering besides these approaches. To address this gap, we therefore propose a novel divisive clustering methodology. Building on the example provided above in Figure~\ref{fig:IntroductionMisspec}, we examine the similarity matrix $\bm{S}$ underlying this example. Specifically, the pairwise distance matrix $\bm{D} = (d_{kl}) = (1 - s_{kl})$ in Figure~\ref{fig:IntroductionMisspec} was derived from the similarity matrix $\bm{S} = (s_{kl})$ containing pairwise similarities of four objects. The corresponding similarity matrix is shown in Figure~\ref{fig:Introduction2}.

Let $\mathcal{S}^p$ be the set of symmetric $p\times p$ matrices. Our goal from the illustrative example above in Figure~\ref{fig:IntroductionMisspec} can be formulated as finding the best block diagonal matrix approximation according to some function 
\begin{equation}\label{eq:MatrixApprox}
   \max\limits_{\bm{B} \in \mathcal{B}_2 } f( \bm{D} - \bm{B} ) \quad \text{or} \quad \min\limits_{\bm{B} \in \mathcal{B}_2 } f(\bm{S} - \bm{B} ) \,,
\end{equation}
where $\mathcal{B}_2 = \{ \bm{B}  : \bm{P}_\pi \bm{B} \bm{P}_\pi^T = \diag(\bm{B}_j, \bm{B}_{j'}) \in \mathcal{S}^{p} = \mathcal{S}^{p_j + p_{j'}}  \}$ is the set of symmetric block diagonal matrices comprising two blocks, and $\bm{P}_\pi$ is a permutation matrix leading to a block diagonal structure. The block diagonal matrix approximation is symmetric since $\bm{D}$ and $\bm{S}$ are symmetric. In the example corresponding to Figure~\ref{fig:IntroductionMisspec}, we thus considered the function 
$$
f(\bm{A}) = \frac{ \Vert  \bm{A} \Vert_{1,1} }{2 p_{j} p_{j'}} \,,
$$
where $\Vert \bm{A} \Vert_{v,v} = \Vert \vectorize(\bm{A}) \Vert_{v} $ denotes the $\mathcal{L}_v$ vector norm for a matrix $\bm{A}$. This approach approximates $\bm{D}$ or $\bm{S}$ by a block diagonal matrix using average linkage, thereby yielding the best split according to the average linkage criterion.

We still have $|\mathcal{B}_2| = 2^{p-1}-1$, meaning there are $2^{p-1}-1$ possible matrix approximations to check. Thus, up to this point, framing clustering as a matrix approximation problem is merely a reformulation. However, in this work, we demonstrate that eigenvectors of the similarity matrix mirror the block diagonal structure of the best matrix approximation according to \eqref{eq:MatrixApprox}. This effectively reduces the number of splits to check. Specifically, the eigenvectors would initially mirror the two blocks $(x_1, x_2, x_3)$ and $x_4$. By calculating the eigenvectors within the first block $(x_1, x_2, x_3)$, they detect the substructure $(x_1, x_2)$ and $x_3$, ultimately leading to the dendrogram as shown in Figure~\ref{fig:IntroductionMisspec} (\textit{right}).

The rest of this article is organized as follows. In Section~\ref{s:HCSVD}, we introduce basic notations and definitions, present the divisive hierarchical clustering using singular vectors methodology based on a similarity matrix, and discuss its extension to distance matrices. Section~\ref{s:Simulation} presents simulation studies, while Section~\ref{s:Examples} illustrates observation and variable clustering on real data and demonstrates how the methodology can accommodate different clustering objectives and construct balanced clusters. Finally, Section~\ref{s:Discussion} concludes with a discussion.

\section{Singular vectors (eigenvectors) for identifying clusters}
\label{s:HCSVD}

\subsection{Setup}

The following notations are used throughout the paper. We use lowercase letters $x$, boldface lowercase letters
$\bm{v}$, boldface capital letters $\bm{X}$, and calligraphic capital letters $\mathcal{G}$ to represent scalars, vectors, matrices, and sets respectively.

Let $\bm{S}$ ($p\times p$) be a matrix of pairwise similarities of $p$ objects $x_1 , \ldots, x_p$ we want to cluster. The singular value decomposition, respectively called eigendecomposition, of this similarity matrix can be written as follows:
\begin{equation*}
    \bm{S} = \bm{V} \bm{\Lambda} \bm{V}^T, \; \bm{V}^T \bm{V} = \bm{I}_p, \; \lambda_1 \geq \cdots \geq \lambda_p \, ,
\end{equation*}
where $\bm{\Lambda}$ contains the singular values (eigenvalues) of $\bm{S}$ on the diagonal, and $\bm{V} = (\bm{v}_1,\ldots,\bm{v}_p)$ are the singular vectors (eigenvectors). Without loss of generality, we assume that all diagonal values of $\bm{S}  = (s_{kl})$ equal one, consequently $|s_{kl}| \leq 1$ because the value one is assigned for the largest similarity, i..e, for the similarity of an object with itself. This can be achieved by scaling the similarity matrix. Similar to the identity of indiscernibles for distance matrix, we further assume that $|s_{kl}| = 1$ if and only if $k = l$. Additionally, we assume that $\bm{S}$ is positive semi-definite, meaning that $\lambda_p \geq 0$.

We will identify clusters according to some clustering objective based on the similarity matrix $\bm{S}$. If in our analysis we have only access to a distance matrix, we can transform this distance to a similarity matrix. This will be discussed in this work and imposes no restriction on the proposed methodology.

In divisive hierarchical clustering, a cluster is recursively split into two disjoint subclusters, starting from the set of all objects. Divisive hierarchical clustering of the objects $x_1,\ldots,x_p$ is therefore a \textit{top-down} approach that can be represented by a sequence of $p$ partitions $\mathcal{P}_{(1)}, \ldots, \mathcal{P}_{(p)}$. Here, $\mathcal{P}_{(1)} = \{ \{ x_1,\ldots,x_p \} \}$ contains all objects as a single cluster (\textit{top}), and $\mathcal{P}_{(p)} = \{ \{x_1\}, \ldots, \{x_p\} \}$ is the partition in which each object forms its separate cluster (\textit{bottom}). Each partition $\mathcal{P}_{(t)}$ is a set of disjoint clusters $\mathcal{G}_{1}, \ldots, \mathcal{G}_{t}$ with $\mathcal{G}_{i} \neq \emptyset$, $\bigcup_{i} \mathcal{G}_{i} = \{ x_1,\ldots,x_p\}$, and $\mathcal{G}_{i} \cap \mathcal{G}_{i'} = \emptyset$ for all $i, i' \in\{1,\ldots,t\}$ and $i \neq i'$. Further, the number of objects contained in each cluster $p_i = |\mathcal{G}_{i}|$ is given by its cardinality. For divisive clustering, $\mathcal{P}_{(t+1)}$ is constructed by splitting one cluster in $\mathcal{P}_{(t)}$. More precisely, at each stage, hierarchical clustering using singular vectors (HC-SVD) splits any cluster $\mathcal{G}_{i}$ contained in the partition $\mathcal{P}_{(t)}$ into two disjoint clusters $\mathcal{G}_{j}$ and $\mathcal{G}_{j'}$ to construct the new partition $\mathcal{P}_{(t+1)}$ that contains $\mathcal{G}_{j}$ and $\mathcal{G}_{j'}$ instead of $\mathcal{G}_{i} = \mathcal{G}_{j} \cup \mathcal{G}_{j'}$. As the clusters are disjoint, it is irrelevant which cluster is selected for the split.

In order to conduct clustering, a measure of similarity or dissimilarity is required. Such measures can originate from different sources, and we present two examples to illustrate the role of the similarity matrix $\bm{S}$.
\begin{enumerate}
    \item \textit{Example 1: Variable clustering}. Consider that $x_1 , \ldots x_p$ are variables. A natural measure that describes similarity between variables is given by the correlation coefficient, thus $\bm{S}$ \textit{can} be the correlation matrix which satisfies positive semi-definiteness and ones on the diagonal. We then split a cluster $\mathcal{G}_{i}$ in such a way that the correlation (similarity) between the variables contained in the two splits $\mathcal{G}_{j}$ and $\mathcal{G}_{j'}$ is minimized according to some clustering objective. 
    \item \textit{Example 2: Observation clustering}. Let $x_1, \ldots, x_p$ now denote observations to be clustered according to a dissimilarity measure, for instance, pairwise Euclidean distances. The distance matrix is transformed into a similarity matrix $\bm{S}$, and a cluster $\mathcal{G}_i$ is then split such that the similarity between the observations in the two resulting clusters $\mathcal{G}_j$ and $\mathcal{G}_{j'}$ is minimized, meaning that the distance between them is maximized.
\end{enumerate}

\subsection{Basic idea}

For now, we only consider splitting one cluster $\mathcal{G}_{i}$ into $\mathcal{G}_{j}$ and $\mathcal{G}_{j'}$ such that $\mathcal{G}_{j} \cup \mathcal{G}_{j'} = \mathcal{G}_{i}$ with $\mathcal{G}_{j} \cap \mathcal{G}_{j'} = \emptyset$. Without loss of generality, we assume $\mathcal{G}_{i} =  \mathcal{P}_{(1)}$, considering the first split in the divisive clustering approach, where all objects $x_1,\ldots,x_p$ are split into two disjoint clusters.

The similarity matrix of can be written as
\begin{equation}\label{eq:ClusterCovarianceMatrix}
    \bm{S} = \begin{pmatrix}
    \bm{S}_j & \bm{S}_{jj'} \\ \bm{S}_{jj'}^T & \bm{S}_{j'}
    \end{pmatrix} \,,
\end{equation}
where $\bm{S}_j$ is the similarity matrix of objects in $\mathcal{G}_{j}$, $\bm{S}_{j'}$ is the similarity matrix of objects in $\mathcal{G}_{j'}$, and $\bm{S}_{jj'}$ contains the pairwise similarities of the objects between these two clusters. We assume this convenient ordering in $\bm{S}$ since this order can be obtained using column permutation.

The objective is to identify the clusters $\mathcal{G}_{j}$ and $\mathcal{G}_{j'}$ in such a way that the similarity between these clusters is minimized, i.e., in such a way that the distance between the objects contained in different clusters is maximized. This corresponds to the best block diagonal matrix approximation $
\min_{\bm{B} \in \bm{B}_2 } f(\bm{S} - \bm{B} )$ outlined in equation \eqref{eq:MatrixApprox} in the introduction.

Let us assume for the moment that $\bm{S}_{jj'} = \bm{0}$. Under this assumption, $\bm{S}$ exhibits a block diagonal structure, meaning that the clusters $\mathcal{G}_{j}$ and $\mathcal{G}_{j'}$ are well separated with similarity values of zero between them. A trivial consequence is that the best block diagonal approximation of $\bm{S}$ is simply $\bm{S}$ itself. \citet{BA25} noted that the eigenvectors of $\bm{S}$ mirror its block diagonal structure.
\begin{Corollary} \label{col:VectorStructure}
The similarity matrix $\bm{S}$ is a block diagonal matrix if and only if the eigenvectors of $\bm{S}$ exhibit the structure
\begin{equation}\label{eq:eigenvectorsMirrorBocks}
    \bm{V} \bm{P}_\pi = \begin{pmatrix}
        \bm{V}_j & \bm{0} \\ \bm{0} & \bm{V}_{j'}
    \end{pmatrix} \, ,
\end{equation}
where $\bm{V}_j$ are the eigenvectors of $\bm{S}_j$, $\bm{V}_{j'}$ are the eigenvectors of $\bm{S}_{j'}$, and $\bm{P}_\pi$ is a column permutation matrix leading to this block diagonal structure.
\end{Corollary}
Therefore, the eigenvectors of $\bm{S}$ mirror the shape of the optimal block diagonal matrix approximation, which identifies the two clusters. Consequently, analyzing the eigenvector structure allows for determining the block diagonal approximation and grouping the objects into distinct clusters $\mathcal{G}_{j}$ and $\mathcal{G}_{j'}$.

However, when $\bm{S}_{jj'} \neq \bm{0}$, the eigenvectors no longer perfectly mirror the best block diagonal approximation, nor the clusters, as they are perturbed by the off-diagonal elements $\bm{S}_{jj'}$ (see \citet{YWS15} alongside an extension in \citet{BD21}). Still, we can use sparse approximations of the eigenvectors to find the optimal block diagonal structure of the similarity matrix compromised by $\bm{S}_{jj'}$ to detect the clusters. \citet{BA25} proposed using sparse eigenvectors, which are approximations of eigenvectors having many zero entries, to detect block diagonal structures in similarity matrices. The key idea is to shrink the zero value components of the eigenvectors perturbed by $\bm{S}_{jj'}$ back to zero, thereby mirroring the block diagonal matrix and clusters yet again.

Following \citet{BA25}, we use the method of \citet{SH08} to obtain sparse eigenvectors based on the following optimization problem:
\begin{equation}\label{eq:OptProblem}
    \min\limits_{\bm{v}} \Vert \bm{S} - \tilde{\bm{v}} \bm{v}^T \Vert_F^2 \;\; {\rm subject \; to} \;\;  \Vert \bm{v} \Vert_2^2 = 1, \; \Vert \bm{v} \Vert_1 \leq c \, .
\end{equation}
This computes the first sparse eigenvector $\check{\bm{v}}_1$ with $\Vert \check{\bm{v}}_1 \Vert_2^2 = 1$ using the $\mathcal{L}_1$ regularization parameter $c$. $\tilde{\bm{v}}$ is allowed to vary and therefore contains the corresponding ``eigenvalue'' $\check{\lambda}$. The remaining sparse eigenvectors $\check{\bm{v}}_l$ for $l>1$ can be calculated iteratively where the correlation matrix $\bm{S}$ must be replaced by the residual matrices of the sequential matrix approximations. We note that the blocks can be revealed using only a few eigenvectors since each block can be mirrored by a single eigenvector. Therefore, it is not necessary to calculate all $p$ sparse eigenvectors. This will be elaborated later in this work.

The steps for splitting any cluster $\mathcal{G}_{i}$, i.e., for identifying the objects of the two diagonal blocks belonging to the splits $\mathcal{G}_{j}$ and $\mathcal{G}_{j'}$, are outlined in Algorithm~\ref{alg:ClusteringSVD}, which illustrates the approach for hierarchical clustering using singular vectors (HC-SVD). Let therefore $s$ be the degree of sparsity for all calculated sparse eigenvectors, i.e. the number of non-zero components in each sparse eigenvector. Similar to (\ref{eq:eigenvectorsMirrorBocks}), each sparse eigenvector mirrors a block diagonal matrix and thus represents a candidate split into two clusters: one cluster is represented by the non-zero components and the other by the components being equal to zero. Therefore, $s$ lies between $1$ and $p_i-1$ where $p_i = |\mathcal{G}_{i}|$ are the number of objects contained in the cluster we want to split, corresponding to number of rows and columns of $\bm{S}_i$.

\begin{Algorithm}\label{alg:ClusteringSVD}
Procedure for splitting any cluster $\mathcal{G}_{i}$ into clusters $\mathcal{G}_{j}$ and $\mathcal{G}_{j'}$, with $\mathcal{G}_{j} \cup \mathcal{G}_{j'} = \mathcal{G}_{i}$ and $\mathcal{G}_{j} \cap \mathcal{G}_{j'} = \emptyset$, i.e., finding the two blocks $\bm{S}_{j}$ and $\bm{S}_{j'}$ that best approximate $\bm{S}_{i}$. \\
    \textbf{Input} Cluster $\mathcal{G}_{i}$ with similarity matrix $\bm{S}_i$ of the objects contained in this cluster. \\
    \textbf{for }$s \in \{ 1, \ldots, p_i - 1 \}$ \textbf{do}
    \begin{enumerate}
        \item Compute the first $q_i$ sparse eigenvectors of the $p_i \times p_i$ similarity matrix $\bm{S}_i$, each with $s$ components equal to zero. Consequently, each sparse eigenvector represents a candidate split, indicated by the zero (one group) and non-zero (second group) values.

        \item Compute the $q_i$ distances between these candidate splits.
    \end{enumerate}
    Among the $q_i(p_i - 1)$ calculated distances, select the clusters $\mathcal{G}_{j}$ and $\mathcal{G}_{j'}$ with the largest distance $f_{jj'} = f(\bm{S}_i - \bm{P}_\pi \diag(\bm{S}_{j}, \bm{S}_{j'}) \bm{P}_\pi^T)$ between them:
\begin{equation*}
    \max\limits_{j,j'} \{ f_{jj'}  :  \mathcal{G}_{j} \cup \mathcal{G}_{j'} =  \mathcal{G}_{i} ;  \mathcal{G}_{j} \cap \mathcal{G}_{j'} = \emptyset \} \,.
\end{equation*} 
    \textbf{Output} The two splits $\mathcal{G}_{j}$ and $\mathcal{G}_{j'}$ of $\mathcal{G}_{i}$.
\end{Algorithm}

To optimally split a cluster $\mathcal{G}_{i}$, we must find $\mathcal{G}_{j}$ and $\mathcal{G}_{j'}$ from $2^{p_i - 1}-1$ candidate partitions of $\mathcal{G}_{i}$, which are too many to be feasible in most applications. However, by using sparse eigenvectors for pre-selecting candidate splits, as in Algorithm~\ref{alg:ClusteringSVD}, this number is reduced to $q_i(p_i - 1)$.

\begin{Remark}\label{re:NoSplits}
For $p_i > 6$, the number of splits to examine for a cluster $\mathcal{G}_{i}$ of $ p_i$ objects is reduced from $2^{p_i - 1}-1$ to $q_i(p_i-1)$, with $q_i \leq p_i$, using sparse eigenvectors as outlined in Algorithm~\ref{alg:ClusteringSVD} .
\end{Remark}

\subsection{Recovery of optimal block diagonal matrix approximation}

In Algorithm~\ref{alg:ClusteringSVD}, all candidate splits are identified by sparse eigenvectors. We now analyze the recovery of eigenvectors that mirror the optimal block diagonal matrix approximation using sparse eigenvectors computed via the optimization problem in \eqref{eq:OptProblem}.

\begin{Theorem}\label{th:RecoveryBound}
Consider a similarity matrix as in \eqref{eq:ClusterCovarianceMatrix} with $\bm{S}_{jj'}  \neq \bm{0}$ ($p_j \times p_{j'}$). The sparse eigenvector $\check{\bm{v}}$ with eigenvalue $\check{\lambda}$ computed using \eqref{eq:OptProblem} mirrors the block diagonal matrix $\diag(\bm{S}_{j} , \bm{S}_{j'})  \in \mathcal{B}_2$ if  
\begin{equation}\label{eq:CondThBound}
\frac{1}{p} \Vert \bm{S}_{jj'} \Vert_{\infty , \infty}\leq \max \{ |\check{\bm{v}}|_{(p_j)} , |\check{\bm{v}}|_{(p_{j'})} \} \,.
\end{equation}
Here, $|\check{\bm{v}}|_{(p)}$ is the $p$-th order statistic of absolute components in $\check{\bm{v}}$.
\end{Theorem}

We now discuss the condition \eqref{eq:CondThBound} in more detail. Single object splits in Algorithm~\ref{alg:ClusteringSVD} are always detected: Any single object split is identified by a standard base vector such as $\check{\bm{v}} = \bm{e}_1$, which corresponds to the case that $\diag(1, \bm{S}_{j'})$ with $p_j = 1$ is the optimal block diagonal approximation. Consequently, $\max \{ |\check{\bm{v}}|_{(p_j)} , |\check{\bm{v}}|_{(p_{j'})} \} = 1$ and $\Vert \bm{S}_{jj'} \Vert_{\infty , \infty} / p <  1$ because $\bm{S}$ is scaled and similarities are therefore not larger than one in absolute terms. This satisfies condition \eqref{eq:CondThBound} in Theorem~\ref{th:RecoveryBound}.

When $p_j, p_{j'} > 1$, however, condition \eqref{eq:CondThBound} becomes more difficult to interpret. It is intuitive that $\Vert \bm{S}_{jj'} \Vert_{\infty , \infty}$ increases with $p_j$ and $p_{j'}$ since $\bm{S}_{jj'}$ then contains more similarity values, although it remains bounded as $\Vert \bm{S}_{jj'} \Vert_{\infty , \infty} <1$. At the same time, $\max\{ |\check{\bm{v}}|_{(p_j)} , |\check{\bm{v}}|_{(p_{j'})} \} $ tends to decrease due to the constraint $\Vert \check{\bm{v}} \Vert_2 = 1 $. As a result, the condition depends on the specific value of these objects, as well as the overall scaling factor $p$ on the left-hand side of the inequality.

Intuitively, the smaller the off-diagonal components in $\bm{S}_{jj'}$, the more likely the inequality in Theorem~\ref{th:RecoveryBound} is satisfied. Therefore, detection of the block diagonal matrix can additionally be increased using $\bm{S}^{\circ h}$ instead of $\bm{S}$. Here, $\odot$ denotes the Hadamard product and $\bm{S}^{\circ h} = \underbrace{ \bm{S} \odot \cdots \odot \bm{S} }_{h \text{ times}}$ such that $\bm{S}^{\circ h} = (s_{kl}^h)$, is the Hadamard power. Using the Hadamard power, most similarities are shrunk towards zero as $|s_{kl}| \leq 1$ while the coefficients in $\bm{S}_{jj'}$ shrink faster towards zero than the coefficients of the blocks $\bm{S}_j$ and $\bm{S}_{j'}$. This allows for better detection according to condition \eqref{eq:CondThBound}. We note that $\bm{S}^{\circ h}$ maintains it definiteness as $\bm{S} \odot \bm{S} \geq {\rm det}(\bm{S}){\rm det}(\bm{S})$ according to the Schur product theorem.

In the supplementary material, we provide a simulation study to assess the robustness of recovering the eigenvectors that mirror the optimal block diagonal approximation. Additionally, we demonstrate that robustness can be increased by taking the Hadamard power $\bm{S}^{\circ h}$ of $\bm{S}$. We note that for too large $h$, all components converge towards zero which becomes numerically unstable. Throughout this work, we will use $h = 2$.

\subsection{Number of sparse eigenvectors}

The use of sparse eigenvectors reduces the number of candidate splits to be examined. We now propose an approach to reduce the number $q_i$ of computed sparse eigenvectors for HC-SVD to improve computation times. To motivate our approach, we again examine the first split of the divisive approach, i.e., we examine where $\mathcal{G}_{i} = \mathcal{P}_{(1)}$ is partitioned into two distinct clusters $\mathcal{G}_{j}$ and $\mathcal{G}_{j'}$.

Consider for now that the similarity matrix $\bm{S}$ of $\mathcal{G}_{i}$ equals a block diagonal matrix with $b$ blocks. Consequently, the eigenvectors $\bm{V} \bm{P}_\pi$ follow the same block diagonal structure when the columns are permuted using a permutation matrix $\bm{P}_\pi$ (see Lemma~1 in \citet{BA25}):
\begin{equation*}
 \bm{S} =   \begin{pmatrix}
    \bm{S}_1 & \bm{0} & \bm{0} \\
    \bm{0} & \ddots & \bm{0}\\
    \bm{0} & \bm{0} & \bm{S}_b
    \end{pmatrix} \,, \qquad 
    \bm{V} \bm{P}_\pi =  \begin{pmatrix}
    \bm{V}_1 & \bm{0} & \bm{0} \\
    \bm{0} & \ddots & \bm{0}\\
    \bm{0} & \bm{0} & \bm{V}_b
    \end{pmatrix} \, .
\end{equation*}
Several optimal block diagonal approximations comprising two blocks exist for this matrix. Let us for now identify the block diagonal matrix $\diag( \bm{S}_1 , \diag(\bm{S}_2 , \ldots, \bm{S}_b) )$ consisting of the two blocks $\bm{S}_1$ and $\diag(\bm{S}_2 , \ldots, \bm{S}_b)$. From a clustering perspective, we therefore consider the split of $\mathcal{G}_{i}$ into the objects associated with the similarity matrix $\bm{S}_1$ to $\mathcal{G}_{j}$, while all remaining objects are assigned to $\mathcal{G}_{j'}$. The first eigenvector of $\bm{S}_1$ contained in $\bm{V}_1$, given as the first column vector of $\bm{V} \bm{P}_\pi$ above, mirrors the block diagonal matrix $\diag( \bm{S}_1 , \diag(\bm{S}_2 , \ldots, \bm{S}_b) )$ and therefore identifies this split.

However, since the permutation matrix $\bm{P}_\pi$ is unknown and eigenvectors are typically ordered by the magnitude of their corresponding eigenvalues, we do not know the index of this first eigenvector of $\bm{V}_1$ that represents the cluster $\mathcal{G}_{j}$ among all eigenvectors. Still, we can assess the minimal number of eigenvectors required to identify each block and therefore also all candidate splits.

\begin{Remark}\label{re:kMin}
    Let the similarity matrix $\bm{S} = \diag(\bm{S}_1 , \ldots, \bm{S}_b)$ have $b$ blocks, ones on the main diagonal, and ordered eigenvalues $\lambda_1 \geq \cdots \geq \lambda_p \geq 0$. If
    \begin{equation*}
        q = \argmin_{l \in \{1,\ldots,p\}} \{ \lambda_l : \lambda_l \geq 1 \} \, ,
    \end{equation*}
    each block $\bm{S}_1 , \ldots, \bm{S}_b$ is reflected by at least one of the eigenvectors $\bm{v}_1, \ldots, \bm{v}_q$. 
\end{Remark}

For any cluster $\mathcal{G}_{i}$ represented by a similarity matrix $\bm{S}_i$, we denote by $q_i$ the number of considered eigenvectors according to Remark~\ref{re:kMin}. In practice, it is unlikely that $\bm{S}$ equals a block diagonal matrix exactly. Still, the intuition that each block is linked to at least one eigenvector with eigenvalue larger or equal to one remains reasonable. Therefore, we suggest to follow the intuition provided in Remark~\ref{re:kMin} and use the number of eigenvalues larger or equal to one as a reference for the number of sparse eigenvectors for HC-SVD. However, due to potential numerical approximations, where eigenvalues close to one might be calculated as slightly smaller than one, we include two additional sparse eigenvectors and compute $\min(q_i+2,p_i)$ sparse eigenvectors in the provided code. There is no strict mathematical reason for choosing exactly two additional eigenvectors.

It is worth noting that even if all $p_i$ sparse eigenvectors are selected for splitting a cluster $\mathcal{G}_{i}$, the number of candidate splits is reduced from $2^{p_i - 1}-1$ to the quadratic number $p_i (p_i - 1)$, which is feasible for data with a moderate number of variables

We require positive semi-definite matrices to support the intuition outlined in Remark~\ref{re:kMin}. Additionally, if the main diagonal of $\bm{S}_i$ contained components larger than one, the number of sparse eigenvectors $q_i$ to examine may increase, making the process computationally more intensive. Conversely, if the main diagonal contains components smaller than one, not all blocks are guaranteed to be represented by at least one eigenvalue larger than one. Later in this work, we discuss how these two assumptions can be implemented in practice.

We note that the number of sparse eigenvectors in Remark~\ref{re:kMin} is equivalent to the Kaiser criterion \citep{KA60}, which can be used to determine the number of factors in factor analysis. Despite this equivalence, the underlying considerations for these two concepts differ: The Kaiser criterion aims to select all eigenvectors with a certain explanatory power, while our criterion selects as many eigenvectors as necessary to ensure a representation of each cluster (block) by at least one eigenvector.

\subsection{Distance for hierarchical clustering}

Clusters are commonly expressed using dissimilarities. While we identify clusters based on a similarity matrix, we now express the results using dissimilarities $\bm{D} = (d_{kl})$. We recap that the optimal splits $\mathcal{G}_{j}$ and $\mathcal{G}_{j'}$ of a cluster $\mathcal{G}_{i}$ are those with the greatest dissimilarity between them. A natural transformation from similarities to dissimilarities is given by $d_{kl} = 1 - | s_{kl} |$ which corresponds to the following linkage functions.

\begin{Theorem}\label{th:InducedDistance}
Consider a similarity matrix $\bm{S}_i = (s_{kl})$ and a dissimilarity matrix $\bm{D}_i = (d_{kl})$ with $d_{kl} = 1 - | s_{kl} |$ of the cluster $\mathcal{G}_{i}$, both denoted according to the disjoint splits $\mathcal{G}_{j}$ and $\mathcal{G}_{j'}$ as in equation (\ref{eq:ClusterCovarianceMatrix}). The linkage function $f_{jj'}$, used to determine the distance between the two cluster $\mathcal{G}_{j}$ and $\mathcal{G}_{j'}$, induced by average or single linkage
\begin{align*}
    &f_{jj'} = \begin{cases} \dfrac{\Vert \bm{D}_{jj'}\Vert_{1,1}}{p_{j} p_{j'}} =  1 - \dfrac{ \Vert  \bm{S}_{jj'} \Vert_{1,1} }{p_j p_{j'}} & j \neq j' \\ 0 & j = j'  \end{cases} \quad \text{(average linkage)} \\
    &f_{jj'} = \Vert \bm{D}_{jj'} \Vert_{\infty, \infty} =  1 - \Vert \bm{S}_{jj'} \Vert_{\infty, \infty} \quad \text{(single linkage)}
\end{align*}
   defines a semidistance (the triangle inequality is not fulfilled).
\end{Theorem}

Using HC-SVD, we can cluster both variables and observations of a data matrix. For clustering variables, linear correlation provides a natural measure of dissimilarity. Specifically, we may consider the similarity matrix $\bm{S} = \bm{R}$ as the matrix of pairwise correlation coefficients $\bm{R} = (r_{kl})$ computed from the data matrix $\bm{X}$. By construction, the correlation matrix is positive semi-definite and has ones on the main diagonal, making it a valid similarity matrix for HC-SVD. Two variables can then considered similar if their absolute correlation is high, and dissimilar if it is low, meaning that $d_{kl} = 1-|r_{kl}|$. In general, other measures of similarity can also be employed as long as the resulting similarity matrix is positive semi-definite with ones on the diagonal.

If we wish to treat negative correlation as dissimilarity—meaning positive correlation corresponds to similarity, no correlation corresponds to medium similarity/dissimilarity, and negative correlation corresponds to dissimilarity—we can instead use the transformed matrix $\bm{R}^+ = 1/2 ( \bm{R} + \bm{1}) $, where $\bm{1}$ denotes the matrix of ones. This transformation maps correlations to the interval $[0, 1]$, with positive correlations mapped to $(1/2, 1]$ and negative correlations to $[0, 1/2)$. This linear transformation ensures that perfectly negatively correlated clusters appear as blocks on the diagonal of $\bm{R}^+$, with zero values between the blocks. Consequently, the intuition of block detection developed earlier in this work, based on Corollary~\ref{col:VectorStructure}, still applies. Importantly, the number of sparse eigenvectors to be examined for HC-SVD, as outlined in Remark~\ref{re:kMin}, remains unchanged because $\bm{R}^+$ is positive semi-definite by construction, since the rank one matrix $\bm{1}$ is positive semi-definite. Additionally, $\bm{R}^+$ retains ones on the diagonal. More generally, this transformation can be applied to any similarity matrix $\bm{S}$, yielding $\bm{S}^+$.

HC-SVD is based on similarity matrices, but it can also be adapted to clustering from a dissimilarity matrix $\bm{D} = (d_{kl})$, which is common in observation clustering, by applying monotone transformations that preserve the relative ordering of the similarities. In this case, an \textit{ad hoc} transformation can be defined as $\bm{S} = \bm{1} - \bm{D}^\ast$, so that $s_{kl} = 1 - d_{kl}^\ast$, where $\bm{D}^\ast = \bm{D} / \Vert\bm{D} \Vert_{\infty, \infty}$. This scaling maps $\bm{D}$ to the unit interval and ensures that all diagonal entries of $\bm{S}$ equal one. However, it does not guarantee that $\bm{S}$ is positive semi-definite.

If the transformation produces a positive semi-definite matrix, HC-SVD can be directly applied to $\bm{S} = \bm{1} - \bm{D}^\ast$, allowing straightforward comparisons between clustering results based on $\bm{S}$ and those obtained from $\bm{D}^\ast$, as similarities and dissimilarities are directly. An example of this approach will be presented later in this work. If positive semi-definiteness is not achieved, alternative transformations can be employed, such as the Gaussian kernel $s_{kl} = \exp( -d_{kl}^2/(2 \gamma^2))$, which ensures positive-definiteness and ones on the diagonal \citep{HSS08}.

\subsection{Hierarchical clustering using singular vectors}

For the complete HC-SVD procedure, it is left to compute a $p \times p$ distance matrix $\bm{U} = (u_{kl})$ that contains the distances between variables within the hierarchical clusters. This matrix is constructed by iteratively applying Algorithm~\ref{alg:ClusteringSVD} to each partition $\mathcal{P}_{(t)}$, as outlined in Algorithm~\ref{alg:HCSVD}.

\begin{Algorithm}\label{alg:HCSVD}
Procedure for HC-SVD on a similarity matrix $\bm{S}$ of objects $x_1 , \ldots, x_p$. \\
    \textbf{Input} Similarity matrix $\bm{S}$ of objects $x_1 , \ldots, x_p$. \\
    Let $\mathcal{P}_{(1)} = \{\{ x_1 , \ldots, x_p \}\}$ be the partition containing all objects as a single cluster, and let $\bm{U} = (u_{kl})$ be a $p \times p$ matrix with zeros on the diagonal which will contain the distances between objects within the hierarchical clusters. \\
    \textbf{for} $t \in \{ 1, \ldots, p - 1 \}$ \textbf{do}
    \begin{enumerate}
        \item Obtain $\mathcal{P}_{(t+1)}$ by splitting any cluster $\mathcal{G}_{i} \in \mathcal{P}_{(t)}$ with similarity matrix $\bm{S}_i$ according to Algorithm~\ref{alg:ClusteringSVD} into $\mathcal{G}_{j}$ and $\mathcal{G}_{j'}$ such that $\mathcal{G}_{j} \cup \mathcal{G}_{j'} = \mathcal{G}_{i}$ with $\mathcal{G}_{j} \cap \mathcal{G}_{j'} = \emptyset$. Use $\bm{S}_i^{\circ h}$ with $h\geq 1$ as input for Algorithm~\ref{alg:ClusteringSVD}.

        \item Compute distances $f_{jj'}$ between the variables of $\mathcal{G}_{j}$ and $\mathcal{G}_{j'}$ and assign them: 
        \begin{equation*}
            u_{kl} = u_{lk} = f_{jj'} \,,
        \end{equation*}
        for all $x_k \in \mathcal{G}_{j}$ and $x_l \in \mathcal{G}_{j'}$.
    \end{enumerate}
    \textbf{Output} Distance matrix $\bm{U}$ of HC-SVD.
\end{Algorithm}

For the very first split when splitting $\mathcal{G}_{i} = \mathcal{P}_{(1)}$ into two clusters $\mathcal{G}_{j}$ and $\mathcal{G}_{j'}$ to obtain the partition $\mathcal{P}_{(2)} = \{\mathcal{G}_{j}, \mathcal{G}_{j'} \}$, the distances between the objects contained in $\mathcal{G}_{j}$ and $\mathcal{G}_{j'}$ are assigned by $f_{jj'}$. In the subsequent step, when the cluster $\mathcal{G}_{j}$ is further split into clusters $\mathcal{G}_{g}$ and $\mathcal{G}_{g'}$, the distances between the objects in $\mathcal{G}_{g}$ and $\mathcal{G}_{g'}$ are updated to $f_{gg'}$, while the distance to the objects in $\mathcal{G}_{j'}$ remain unchanged. Thus, the dissimilarity between variables is refined iteratively, identifying the hierarchical structure as the algorithm progresses.

It is desirable that the resulting distance matrix is an ultrametric distance matrix. This property ensures that the distance between the clusters decreases with each split, making the resulting dendrogram interpretable without inversions \citep{JO67, HA67}. A distance matrix $\bm{U} = (u_{kl})$ is ultrametric if
\begin{equation*}
    u_{kl} \leq \max\{ u_{km} , u_{ml} \} \, ,
\end{equation*}
for all $k$, $l$, and $m$.

\begin{Theorem}\label{th:UltrametricDistanceMatrix}
The distance matrix $\bm{U}$ obtained by HC-SVD using distances $f_{jj'}$ induced by average linkage or single linkage as described in Theorem~\ref{th:InducedDistance} is ultrametric if condition \eqref{eq:CondThBound} in Theorem~\ref{th:RecoveryBound} holds for $\bm{S}_i^{\circ h}$ in each iteration of HC-SVD.
\end{Theorem}
We now provide an intuition of the role of Theorem~\ref{th:RecoveryBound} in HC-SVD. In \eqref{eq:CondThBound}, we discussed below Theorem~\ref{th:RecoveryBound} that  $\Vert \bm{S}_{jj'} \Vert_{\infty,\infty}$ tends to increase with $p_j$ and $p_{j'}$, while $\max\{ |\check{\bm{v}}|_{(p_j)} , |\check{\bm{v}}|_{(p_{j'})} \} $ decreases. Intuitively, at the beginning of divisive clustering, $\max\{ |\check{\bm{v}}|_{(p_j)} , |\check{\bm{v}}|_{(p_{j'})} \} $ is small relative to the end of the procedure. At the same time, clusters are farther apart, thus $\Vert \bm{S}_{jj'} \Vert_{\infty,\infty}$ is smaller compared to later stages of clustering. However, $p = p_j + p_{j'}$ is large, meaning that $1/p$ is small. As the hierarchical clustering progresses, the groups become smaller and closer to one another. This leads to larger values of $\Vert \bm{S}_{jj'} \Vert_{\infty,\infty}$ ,and $1/p$ increases because $p$ decreases. Unlike the initial stages, however, $\max\{ |\check{\bm{v}}|_{(p_j)} , |\check{\bm{v}}|_{(p_{j'})} \} $ also increases because $p_j$ and $p_{j'}$ are smaller.

\section{Simulation study}
\label{s:Simulation}

\subsection{Preliminaries}
In this section, we evaluate the performance of HC-SVD as described in Algorithm~\ref{alg:HCSVD}. We compare HC-SVD to five different methods: Three of them, DIANA \citep{KR90}, ACCA \citep{BD08, BD10}, and UCOR \citet{CVZ20, CVZ24} were discussed in the introduction in Section~\ref{s:Introduction}. Additionally, we consider $k$-medoids (partitioning around medoids, PAM), and spectral clustering with the ratio cut (Rcut) and the normalized cut (Ncut) objective \citep{KR90, vL07}. Although these two methods do not inherently create a hierarchical structure, they can be applied directly to distance matrices and can be used to compute $2$ to $p-1$ clusters, mimicking hierarchical clustering.

All computational results were conducted in \texttt{R} 4.1.3 \citep{RSoftware} on a PC running macOS Sequoia 15.6.1. with $8$ GB of RAM. Code to replicate the simulations alongside details about computational implementations for all methods (B Online Appendix) are provided in the supplementary material. Furthermore, HC-SVD has been implemented specifically for this research project in the \texttt{R} package \texttt{blox} \citep{BA25Rpackage}.

\subsection{Simulation study designs}

In design A, we generate dendrograms by randomly splitting $p$ variables for $100$ simulation iterations, yielding $100(p-1)$ clusters. The ultrametric distance matrix $\bm{U}=(u_{kl})$ generated by the dendrogram is mapped to the unit interval via $\sigma_{kl} = \exp(-u_{kl}) $, resulting in the population similarity (correlation) matrix $\bm{\Sigma} = (\sigma_{kl})$ of the constructed hierarchical variable structure. We then simulate a data matrix $\bm{X}$ ($n\times p$) with correlation matrix $\bm{\Sigma}$ and compute the sample correlation matrix as similarity matrix $\bm{S}$. In this design, we set $ n = 2p$ and $p \in \{100, 200, 300, 400, 500\} $.

Clustering methods differ in their objectives. Because design A is based on an ultrametric structure, the population clusters are recognized identically across methods, allowing a direct comparison. However, ultrametricity also makes clusters easier to detect, motivating the use of an alternative study design.

In design B, we generate random correlation matrices without imposing ultrametricity. Different methods may thus construct different clusters depending on they objective, meaning that we cannot directly compare the methods. Instead, each method is evaluated relative to its own true cluster structure, identified via exhaustive search. We then assess whether the method recovers this structure and compare HC-SVD against the other methods accordingly.

To generate the similarity matrix, we first construct a symmetric positive semi-definite matrix $\bm{F}\bm{F}^T$, where $\bm{F}$ ($p\times p$) has independent entries drawn from $U(-0.4,1)$. We then scale $\bm{F}\bm{F}^T$ to obtain matrix $\bm{S}$. Because exhaustive search is computationally demanding, we set $p=25$ and run $S=1000$ iterations. We restrict the simulation to finding two clusters (the first split in divisive clustering).

In design A, the number of sparse eigenvectors used for HC-SVD is chosen according to Remark~\ref{re:kMin} due to the large number of variables, whereas in design B all sparse eigenvectors are considered. Results for HC-SVD using all sparse eigenvectors in design A are provided in the supplementary material (Online Appendix C).

For both designs, we evaluate clustering performance using the adjusted Rand index (ARI; \citet{RA71, HA85}), which quantifies similarity between clusters. Positive values indicate a strong match, an ARI of one means a perfect match, an ARI of zero indicates that the clustering is no better than random clustering, and negative values mean poorer performance than random clustering.

\subsection{Simulation results}

\begin{figure}[!htbp]
\center
\includegraphics[width=\textwidth]{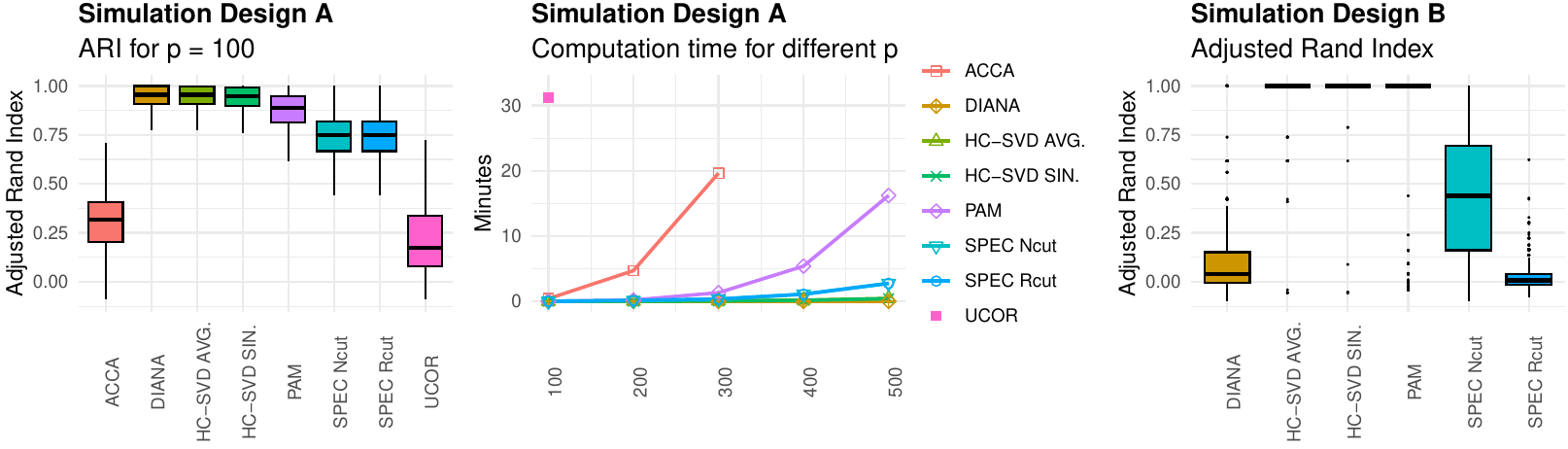}
  \caption{\textbf{Design A}: Performance of ACCA, DIANA, HC-SVD (with both average and single linkage), PAM, spectral clustering (with both ratio cut (SPEC Rcut) and normalized cut (SPEC Ncut) objective), and UCOR, for simulation design A with $p = 100$, measured by ARI (\textit{left}). Performance is evaluated across all possible splits $k\in\{2,\ldots,99\}$. Average computation times (in minutes) are shown for all methods (\textit{center}). Results for $p \in \{200, 300, 400, 500\}$ show similar ARI performance and are provided in the supplementary material, alongside visualizations of the generated dendrograms and computation times separately for the five fastest methods to facilitate comparison.\newline
  \textbf{Design B}: Performance of DIANA, HC-SVD (with both average and single linkage), PAM, spectral clustering (with both with the ratio cut (SPEC Rcut) and the normalized cut (SPEC Ncut) objective), and UCOR, measured by ARI (\textit{right}).}
  \label{fig:ResultSimulation}
\end{figure}

The results are illustrated in Figure~\ref{fig:ResultSimulation} for both design A (\textit{left}) and design B (\textit{right}), alongside the computation times for design A (\textit{center}). ACCA and UCOR are only computed for $p \in \{100, 200, 300\}$ and $p = 100$, respectively, due to their long runtimes of 20 minutes per iteration for ACCA and over 30 minutes for UCOR. They are also not included in design B.

DIANA, HC-SVD and SPEC exhibit runtimes that we would describe as practicable for applied use. For example, DIANA and HC-SVD complete in under 30 seconds (0.5 minutes) on average when $p\leq 500$. However, it is important to note that the computation time for HC-SVD is influenced by the size of the clusters during the initial splitting steps: if many small clusters are identified at the beginning, the number $p_i$ in subsequent steps does not decrease much relative to $p$.

Both design A and B are based on randomly generated matrices. DIANA and HC-SVD perform similarly in design A. Thus, DIANA appears to identify the ultrametric structure reliably. In the more complex design B, where the underlying distance matrix does not exhibit the ultrametric property, HC-SVD and PAM show good performance in matching their true cluster objective. Due to the exhaustive search in design B, we must restrict the simulation study to $p=25$ for computational reasons. It thus remains unclear if PAM performs worse for larger $p$, as it is indicated in design A.

In the supplementary material, we also evaluate the performance of agglomerative clustering with average and single linkage for design A. However, as outlined in the introduction, agglomerative clustering has inherent drawbacks due to its bottom-up construction. This is further elaborated in Section~\ref{s:Examples}.

\subsection{Comments about Principal Component Analysis}

An intuitive question is whether principal component analysis (PCA) can be used to identify variable clusters, given its natural connection to HC-SVD. However, \citet{BA25} demonstrated that PCA already performs poorly in detecting uncorrelated variable clusters, making it unsuitable for hierarchical variable clustering. A detailed discussion of this limitation is provided in the referenced work. Additionally, we illustrate the infeasibility of PCA for variable clustering in the supplementary material (E Online Appendix).

\section{Examples}\label{s:Examples}

\subsection{Preliminaries}
All computational results were conducted in \texttt{R} 4.1.3 \citep{RSoftware} on a PC running macOS Sequoia 15.6.1. with $8$ GB of RAM, and \texttt{R} code to replicate all examples is available in the supplementary material. For HC-SVD, we compute all sparse eigenvalues due to the moderate number of variables.

\subsection{Crime rate by US state}
We illustrate HC-SVD using the well-known \texttt{USArrests} data set, which reports statistics on arrests per $100\,000$ residents for assault, murder, and rape in the 50 US states in 1973, along with the percentage of the population living in urban areas. Originally published in the 96th Annual Edition of the Statistical Abstract of the United States, this data set has since become a classical example in textbooks such as \citet{JWTT21} and is also included as a demonstration data set for the agglomerative clustering function \texttt{hclust(...)} in \texttt{R}. Consequently, it has been extensively studied in the context of agglomerative clustering.

\begin{figure}[!htbp]
\center
\includegraphics[width=\textwidth]{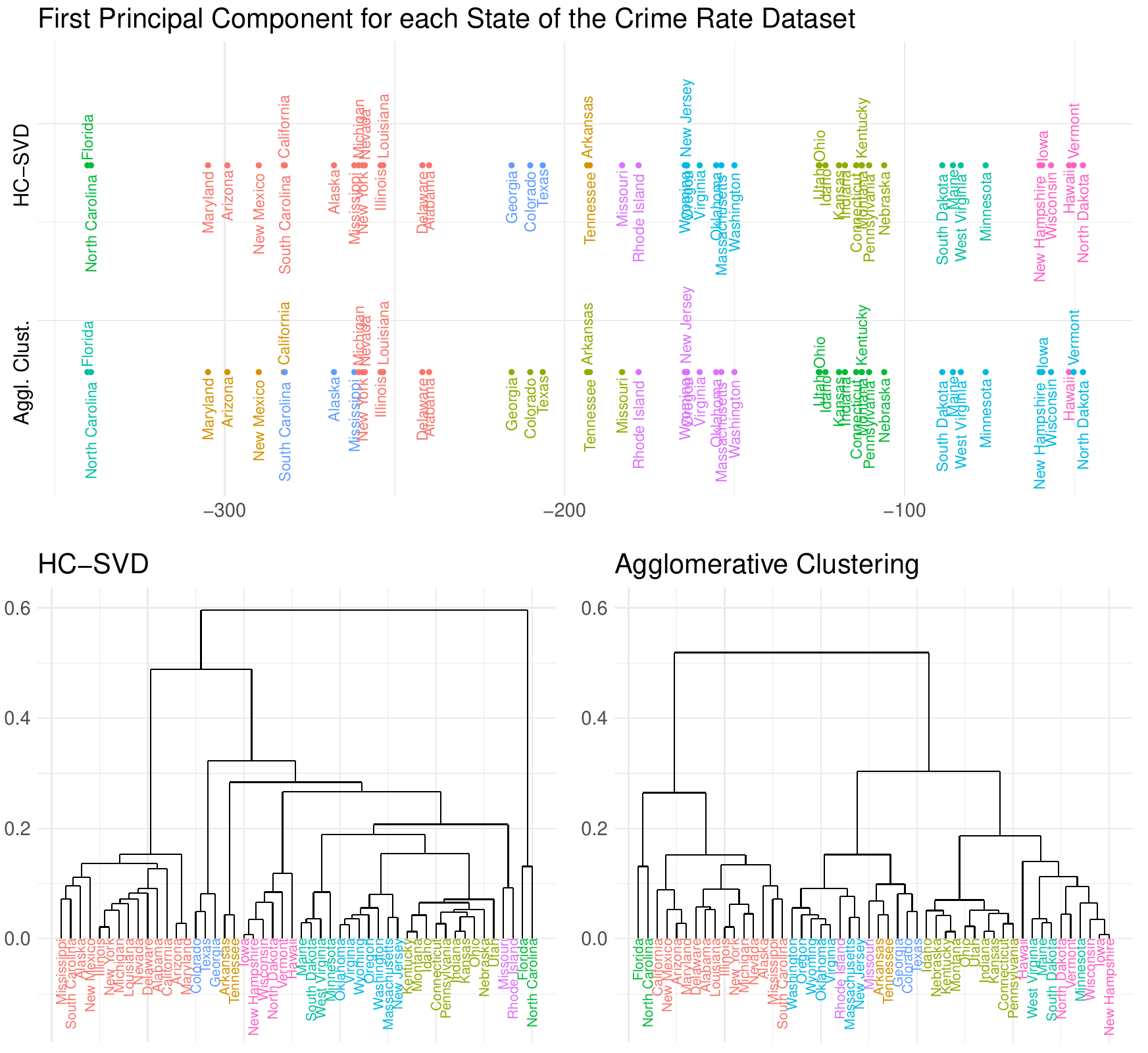}
  \caption{Results for HC-SVD and agglomerative clustering, both using average linkage. \textbf{Upper Figure}: One-dimensional approximation of the crime rate dataset using PCA, where the states are arranged horizontally based on their approximated principal component values. At the \textit{top}, observations are color-coded to represent the first nine clusters identified by HC-SVD. At the \textit{bottom}, observations are color-coded to reflect the clusters identified by agglomerative clustering, highlighting the differences between the two methods. \textbf{Lower Figure:} Dendrograms identified by HC-SVD (\textit{left}) and agglomerative clustering (\textit{right}). The states in \textit{both} dendrograms are color-coded according to the first nine clusters identified by HC-SVD to highlight the differences between the two methods.}
  \label{fig:USArrests}
\end{figure}

Despite this, agglomerative clustering with average linkage fails to identify the clusters that are most distinct. For this analysis, we compute agglomerative clustering using the scaled euclidean distance matrix $\bm{D}^\ast$, while for HC-SVD we use the similarity matrix $\bm{S}$ obtained as $s_{kl} = 1 - d^\ast_{kl}$, which is positive semi-definite for this data set. Figure~\ref{fig:USArrests} illustrates the results for HC-SVD (\textit{left}) and agglomerative clustering (\textit{right}) both using average linkage. Additionally, we project the data to the one dimensional space using PCA in Figure~\ref{fig:USArrests} (\textit{top}) for better visualization. The relative approximation error of a rank one approximation in the Frobenius norm is $13.96\%$ for this data set. Thus, the first principal component gives a reasonable representation.

The initial split into two clusters differs between HC-SVD and agglomerative clustering. HC-SVD identifies clusters with a distance of $0.60$, whereas agglomerative clustering finds a smaller distance of $0.52$, thus failing to detect the most distinct separation. As shown in Figure~\ref{fig:USArrests} (\textit{top}), HC-SVD generally identifies clearer groups, while agglomerative clustering produces more cluttered partitions.

As mentioned in the introduction, this is due to the greedy nature of the ``bottom-up'' agglomeration methods, which we will now discuss in more detail. At the lower levels of the dendrogram, Rhode Island (\textit{lila}) is clustered with Massachusetts and New Jersey (\textit{turquoise}), despite a larger distance being achieved if Massachusetts and New Jersey were first grouped with Oklahoma, Oregon, Virginia, Washington, and Wyoming (all \textit{turquoise}), followed by clustering with Rhode Island.  This misstep in the clustering process persists at higher levels and ultimately prevents the identification of the clusters with the largest distances at the top.

\begin{figure}[!htbp]
\center
\includegraphics[width=\textwidth]{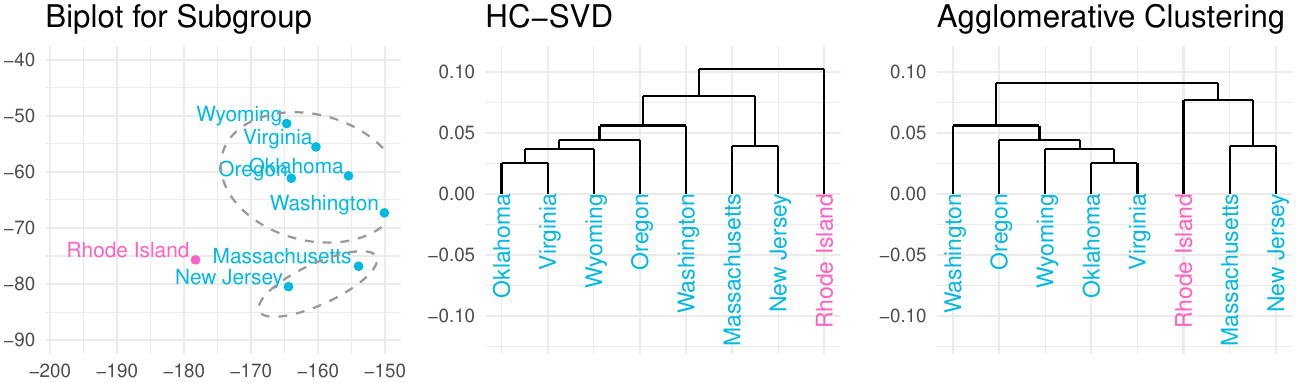}
  \caption{This figure examines the subgroup consisting of nine states: Massachusetts, New Jersey, Oklahoma, Oregon, Virginia, Washington, Wyoming (all \textit{turquoise}), and Rhode Island (\textit{lila}). The figure displays a two-dimensional approximation using PCA (\textit{left}), referred to as biplot, alongside the dendrograms identified by HC-SVD (\textit{center}) and agglomerative clustering (\textit{right}), with average linkage used in both cases.}
  \label{fig:USArrests2}
\end{figure}

This specific case is illustrated in Figure~\ref{fig:USArrests2}, alongside a two-dimensional scatterplot approximation of the nine states using PCA (\textit{left}), referred to as biplot. The relative approximation error of this rank two approximation in the Frobenius norm is $3.43\%$, providing a reasonable visualization of the observation structure within this subgroup. Following the first five agglomerative clustering steps for this subgroup, two clusters emerge: $\mathcal{G}_{j} = \{ \text{Massachusetts}, \allowbreak \text{New Jersey} \} $ (\textit{bottom-right} in the biplot) and  $\mathcal{G}_{j'} = \{ \text{Virginia}, \allowbreak \text{Oklahoma}, \allowbreak \text{Wyoming}, \allowbreak \text{Oregon}, \allowbreak \text{Washington} \} $ (\textit{top-right}  in the biplot). Now, in the subsequent step for the agglomerative approach, where only these two formed clusters $\mathcal{G}_{j}$ and $\mathcal{G}_{j'}$ are considered instead of having the divisive perspective from the top, Rhode Island is closer to $\mathcal{G}_{j}$ (distance $0.04287$) than $\mathcal{G}_{j'}$ is to $\mathcal{G}_{j}$ (distance $0.05278$). This results in the observed agglomerative clustering structure.

\subsection{Big Five personality traits}
In this section, we demonstrate the application of HC-SVD on personality traits. Specifically, we analyze a data set comprising 25 personality items, categorized into five groups: neuroticism (N), extraversion (E), openness to experience (O), agreeableness (A), and conscientiousness (C), according to the NEO personality inventory for personality psychology \citep{CM78, MCM05}. These personality traits align with the Big Five personality traits \citep{RSSK02}. The data set contains self-assessments from 190 university students who indicate the extent to which they embody the personalities described by the 25 items. Data was collected by Adachi \citep{AT18} and can be accessed online from Osaka University at \href{http://bstat.jp/en_material/}{http://bstat.jp/en\_material/}.

\begin{figure}[!htbp]
\center
\includegraphics[width=1.1\textwidth]{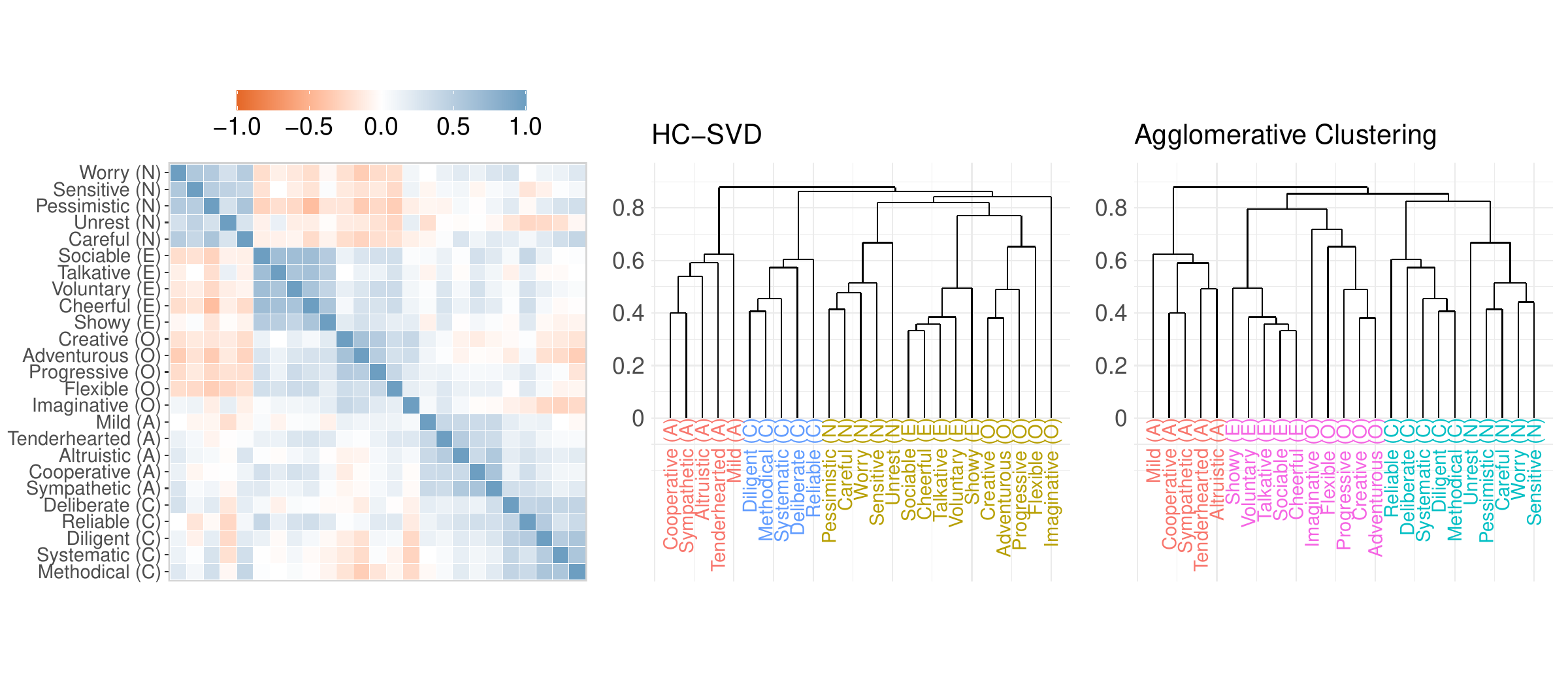}
  \caption{Sample correlation matrix of the Big Five personality traits data \textit{(left)}, along with the hierarchical cluster structure identified by HC-SVD \textit{(center)} and agglomerative clustering \textit{(right)} using average linkage. In the dendrograms, the first split is red coded and the subsequent splits are colored blue and yellow, and pink and turquoise respectively. For improved readability, enlarged versions of these plots are provided in the supplementary material.}
  \label{fig:BigFiveResult}
\end{figure}

In social and behavioral science, internal consistency of item sets is often assessed based on correlations withing these item sets, closely related to average correlation (see, e.g., \citet{ZRYL05} for an overview). Thus, average linkage clustering using the correlation matrix as similarity matrix is appropriate for this example. Figure~\ref{fig:BigFiveResult} shows the sample correlation matrix $\bm{R}$ of the data \textit{(left)} alongside the hierarchical cluster structures identified by HC-SVD \textit{(center)} using average linkage on the similarity matrix $\bm{R}$, and agglomerative clustering \textit{(right)} using average linkage on the distance matrix $\bm{D}  = d_{kl}$ with $d_{kl} = 1 - | r_{kl}|$. The accompanying dendrograms illustrate that both HC-SVD and the agglomerative approach identify the variables corresponding to agreeableness (A), highlighted in red, as the one with the greatest distance from the other variables. However, the subsequent split differs between the two methods. HC-SVD identifies a split with a distance of $0.86$ (colored blue and yellow), which exceeds the $0.85$ distance separating the clusters found by the agglomerative approach (colored pink and turquoise). Thus, based on the sample correlation matrix, agglomerative clustering fails to detect the three clusters with the largest distance between them, similarly to our illustration earlier in the introduction.

\subsection{Adapting different clustering objectives}

HC-SVD is flexible by construction: it can adapt to different clustering objectives by choosing appropriate linkage functions. However, we recall that HC-SVD only considers candidate splits into clusters $\mathcal{G}_{j}$ and $\mathcal{G}_{j'}$ where the between-cluster similarity $\bm{S}_{jj'}$ is ``small'', which restricts it to linkage functions consistent with the bound in Theorem~\ref{th:RecoveryBound}.

DIANA minimizes between-cluster similarity (or maximizes dissimilarity), which aligns with HC-SVD using average linkage. The Rcut objective in spectral clustering is a weighted version of average linkage. Therefore, while not identical, this objective on $\bm{S}_{jj'}$ is still similar to HC-SVD. For Ncut the weighted transformation to $\bm{S}_{jj'}$ differs substantially, making its clusters less likely to be recovered by HC-SVD. A detailed discussion is provided in the supplementary material (B Online Appendix). Nevertheless, we equip HC-SVD with these objectives to evaluate performance. In contrast, PAM optimizes within-cluster similarity in $\bm{S}_{j}$ and $\bm{S}_{j'}$ relative to a centroid, which differs fundamentally from HC-SVD and thus cannot be replicated.

\begin{figure}[!htbp]
\center
\includegraphics[width=0.8\textwidth]{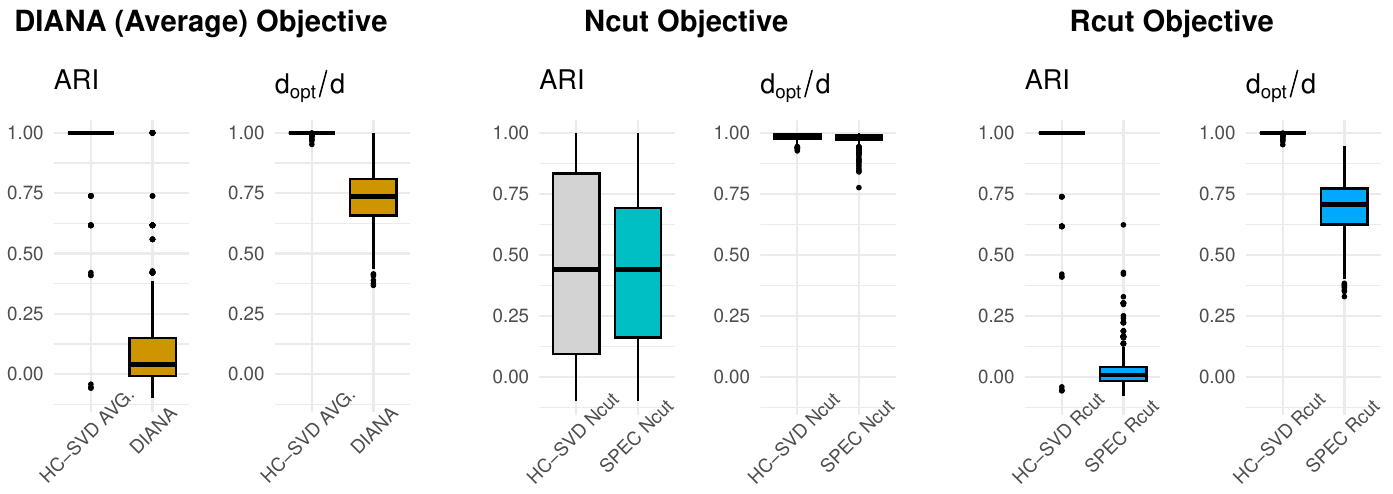}
  \caption{Performance of HC-SVD with clustering objectives from DIANA, spectral clustering with Ncut (SPEC Ncut) and Rcut (SPEC Rcut) in design B. Performance is evaluated using ARI and the ratio $d_{\text{opt}}/d$ between the optimal and detected distance.}
  \label{fig:SimulationMim}
\end{figure}

We replicate design B from Section~\ref{s:Simulation}, applying HC-SVD with the objectives of DIANA and spectral clustering. Performance is evaluated using ARI, and we also compute the ratio of the optimal distance found by exhaustive search to the distance detected by each method, $d_{\text{opt}}/d$. Ratios closer to one indicate better recovery, with equality meaning the true structure was found. The results are shown in Figure~\ref{fig:SimulationMim}.

As discussed above, besides average linkage (DIANA’s objective), HC-SVD can also adapt the Rcut objective, because it relates to the bound established in Theorem~\ref{th:RecoveryBound}. Even though the Ncut objective differs more strongly from the bound in Theorem~\ref{th:RecoveryBound}, HC-SVD performs similarly to spectral clustering in this case. While ARI scores for Ncut drop, the ratio $d_{\text{opt}}/d$ remains close to one. This indicates that although the true clusters are not recovered, the identified clusters have nearly maximal separation under this objective.

\subsection{Balanced clustering}
Balanced clustering, introduced to address skewed cluster sizes, has received increasing attention (see, e.g., \citet{ZNL17}, \cite{ZNWWL24}, and \citet{NXWL25}). HC-SVD is naturally suited for balanced clustering, as cluster sizes can be controlled by constraining the sparsity of the singular vectors.

We recall from Algorithm~\ref{alg:ClusteringSVD} that for a similarity matrix $\bm{S}_i$ ($p_i \times p_i$), we compute all sparsity levels $s \in \{1,\ldots,p_i\}$ to identify candidate splits. If $p_i$ is even and the goal is a perfectly balanced partition, i.e., a $50{:}50$ split, we can restrict to $s=p_i/2$, thereby considering only even-sized splits. For an odd number such as $p_i + 1$, we can restrict $s\in\{ p_i/2 , p_i/2 + 1\}$. More generally, desired balance levels such as $30{:}70$ or $20{:}80$ can be enforced by adjusting the allowed sparsity levels accordingly.

As an illustration, we consider the chaining effect often observed with single linkage clustering. We use the classical \texttt{iris} dataset, which is available and described in \texttt{R} and contains measurements on 150 flowers from three species (setosa, versicolor, and virginica). We use the scaled euclidean distance matrix $\bm{D}^\ast$ and the similarity matrix $\bm{S}$ obtained as $s_{kl} = 1 - d^\ast_{kl}$, which is positive semi-definite for this data set. Agglomerative hierarchical clustering and HC-SVD, both using single linkage, perfectly identifies setosa. We therefore focus on the remaining two species, versicolor (\textit{blue}) and virginica (\textit{orange}).

\begin{figure}[!htbp]
\center
\includegraphics[width=\textwidth]{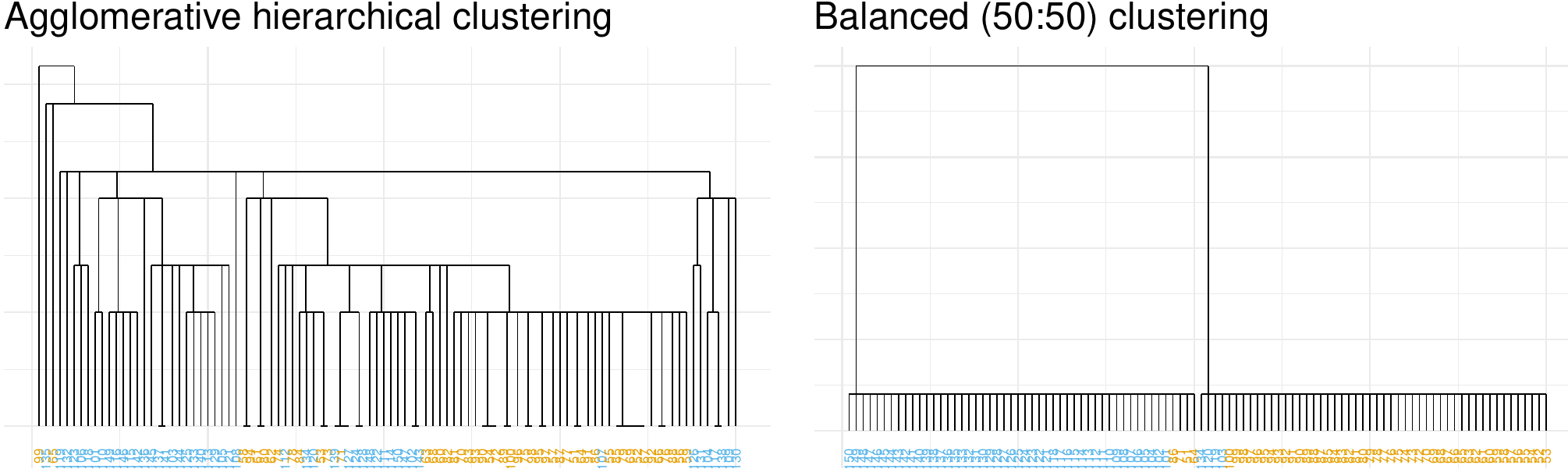}
  \caption{Dendrogram from agglomerative hierarchical clustering with single linkage on the iris dataset (\textit{left}) containing versicolor (\textit{blue}) and virginica (\textit{orange}), and the first split from HC-SVD with single linkage under a balanced $50{:}50$ constraint (\textit{right}).}
  \label{fig:iris}
\end{figure}

Figure~\ref{fig:iris} (\textit{left}) shows the dendrogram identified with agglomerative hierarchical clustering using single linkage. The \textit{right} panel shows the first split identified by HC-SVD using single linkage with enforced equal-sized clusters and distinguishes most of the observations. Since balancing constraints violate ultrametricity, only the first split is illustrated.

\subsection{Revelle's beta}

In psychometrics, \citet{RE79} introduced \textit{Revelle's beta}, a reliability measure based on the worst split-half of a test or scale, to assess general factor saturation. Yet for more than 45 years, no reliable method has been available to compute it, as existing approaches are either computationally infeasible or insufficiently accurate in identifying the worst split-half. HC-SVD overcomes this limitation, as the reliability measure is conceptually related to divisive hierarchical clustering with average linkage. A detailed discussion of this approach and its extension to identify multidimensional structures in tests is provided in \citet{BA25Beta}.

\section{Discussion}
\label{s:Discussion}

In this paper, we introduced a methodology for divisive hierarchical clustering. Our approach uses sparse approximations of the eigenvectors to effectively reduce the number of candidate splits in a divisive approach, enabling fast computation, as demonstrated through simulation studies.

We adapted average and single linkage to divisive clustering and showed that the resulting  HC-SVD distance matrix, constructed using these linkage functions, satisfies the ultrametric property. This ensures that the resulting dendrogram is interpretable without inversions.

Through two empirical examples, we demonstrated the applicability of HC-SVD for observation and variable clustering using real data, and highlighted its advantage over agglomerative clustering, which is commonly used for hierarchical clustering.

We also explored three extensions: adapting objectives from other clustering methods, balanced clustering, and computing reliability measures in psychometrics. In an illustrative simulation study, HC-SVD indicated to outperform the original clustering methods, though theoretical confirmation in line with Theorem~\ref{th:RecoveryBound} and more extensive simulations remain necessary. The same holds for balanced clustering, where HC-SVD provides a promising starting point. Finally, HC-SVD enables the computation of Revelle’s beta, a psychometric reliability measure that had remained practically inaccessible for over 45 years. Extending HC-SVD to other divisive reliability measures is a natural next step.

A potential limitation of HC-SVD arises in spiked correlation scenarios, where the eigenvalues are concentrated in a bulk. In such cases, the number of sparse eigenvectors needed might not be reliably given by the intuition provided in Remark~\ref{re:kMin}, and a larger number of sparse eigenvectors might be required. This, in turn, increases the computation time of HC-SVD. Furthermore, as implied by Theorem~\ref{th:RecoveryBound}, HC-SVD may struggle to detect the hierarchical clustering structure when all variables are strongly correlated, that is when all elements exhibit high similarity.

Possible extensions are the incorporation of linkage functions designed specifically for divisive clustering, offering a novel perspective on clustering methodologies. One natural linkage would be based on the RV coefficient \citep{RE76} which is a scaled version of average linkage. For completeness, the RV coefficient is implemented in the code of the simulation study in the supplementary material and demonstrates good performance. As another extension, measures like the distance correlation \citet{SRB07} for variable clustering could be explored, as they allow the calculation of both distances between groups for the \textit{top} and pairwise distances for the \textit{bottom}, something that cannot be implemented using agglomerative clustering.

HC-SVD is available in the \texttt{R} package \texttt{blox} \citep{BA25Rpackage}.

\section*{Supplementary Material}

\begin{description}
\item[Online Appendix:] This online appendix contains derivations and proofs for the results presented in this paper together with other supplementary material for the simulation studies (for Section~\ref{s:Simulation} and for Theorem~\ref{th:RecoveryBound}), and computational details of this work (OnlineAppendix.pdf).
\item[Examples:] \texttt{R} code to replicate the examples from Section~\ref{s:Examples}.
\item[Simulation Studies:] \texttt{R} code to replicate the simulation studies from Section~\ref{s:Simulation} and regarding Theorem~\ref{th:RecoveryBound} (the corresponding simulation study is available in the online appendix).
\end{description}
Supplementary data are available are available upon request.

\section*{Acknowledgements} I am grateful to Carlo Cavicchia for providing \texttt{R} code for the ultrametric correlation model during early stages of this project.

\nocite{HJ13}
\nocite{KA04}
\nocite{KSH22}
\nocite{LSHM10}
\nocite{MRSHH22}
\nocite{TI96}
\nocite{WTH09}
\nocite{SS25}
\nocite{ZCB24}

\bibliographystyle{chicago}

\bibliography{Bibliography}
\end{document}